\documentclass[aps,twocolumn,pra,superscriptaddress,amsmath,amssymb]{revtex4-1}
\newcommand{\bras}[1]{\langle#1\rvert}
\newcommand{\kets}[1]{\lvert#1\rangle}

\newcommand{\means}[1]{\langle#1\rangle}

\ifx\pdfoutput\undefined
\usepackage[dvipdfmx]{graphicx}
\usepackage[dvipdfmx]{hyperref}
\else
\usepackage{graphicx}
\usepackage{hyperref}
\fi

\usepackage{txfonts}
\usepackage[T1]{fontenc}
\usepackage{xspace}

\usepackage{dcolumn}
\usepackage{bm}
\usepackage{amsmath}
\usepackage{booktabs}
\usepackage[usenames]{color}
\usepackage{xcolor}

\usepackage{ulem}

\begin{document}
\let\emph\textit

\title{
Interlayer Coupling Effect on a Bilayer Kitaev Model
}

\author{Hiroyuki Tomishige}
\affiliation{Department of Physics, Tokyo Institute of Technology, Meguro, Tokyo 152-8551, Japan}
\author{Joji Nasu}
\affiliation{Department of Physics, Tokyo Institute of Technology, Meguro, Tokyo 152-8551, Japan}
\author{Akihisa Koga}
\affiliation{Department of Physics, Tokyo Institute of Technology, Meguro, Tokyo 152-8551, Japan}

 \date{\today}
\begin{abstract}
We investigate a bilayer Kitaev model, where two honeycomb layers are coupled by the Heisenberg interactions, to discuss effects of an interlayer coupling against the Kitaev quantum spin liquids (QSLs).
In this model, there exists a local conserved quantity, which results in no long-range spin correlations in the system.
Using the exact diagonalization, bond operator mean-field theory, and cluster expansion techniques, we study ground state properties in the system.
The obtained results suggest the existence of a first-order quantum phase transition between the Kitaev QSL and singlet-dimer states.
We find that one-triplet excitation from the singlet-dimer ground state is localized owing to the existence of the local conserved quantity.
To examine finite-temperature properties, we make use of the thermal pure quantum state approach.
We clarify that double-peak structure in the specific heat inherent in the Kitaev QSL
is maintained even above the quantum phase transition.
The present results suggest that the Kitaev QSL is stable against the interlayer interference.
Magnetic properties of multilayer Kitaev models are also addressed.
\end{abstract}

\maketitle
\section{Introduction}

Exploring quantum spin liquids (QSLs) is one of the central subjects in condensed matter physics since the Anderson's suggestion~\cite{Anderson1973,Balents2010,Zhou2017,Savary2017}.
A lot of theoretical and experimental studies have been devoted to clarify the nature of QSLs but the thermodynamic properties and excitation spectra remain elusive.
Theoretically, it has been still difficult to analyze the properties of frustrated Heisenberg models, which are considered to be archetypal models of QSLs, without approximations.
On the other hand, one of the promising models to discuss QSLs is a quantum spin model on a honeycomb lattice with bond-dependent Ising interactions, which is known as the Kitaev model~\cite{Kitaev2006,Trebst2017pre,Hermanns2017pre}.
This model is exactly solvable and its ground state is a QSL
with short-range spin correlations~\cite{Baskaran2007,Knolle2014}.
Furthermore, quantum spins are fractionalized into
itinerant Majorana fermions and localized $Z_2$ fluxes.
This leads to gapless elementary excitations in the ground state and
double peak structure in the specific heat~\cite{Kitaev2006,Feng2007,Chen2007,Chen2008,Pedrocchi2011,Nasu2014,Nasu2015}.
Moreover, Jackeli and Khaliullin have revealed that
the Kitaev model should be realizable in Mott insulators
with the strong spin-orbit coupling and specific lattice structure,
where localized  $j_{\rm eff}=1/2$ spins are coupled
by the superexchange interactions~\cite{Jackeli2009}.
As the candidate materials, $A_2$IrO$_3$($A$=Na, Li)~\cite{Singh2010,Singh2012,Comin2012,Choi2012},
$\alpha$-RuCl$_3$~\cite{Plumb2014,Kubota2015,Sears2015,Majumder2015} and $\rm H_3LiIr_2O_6$~\cite{Bette2017} have been intensively studied,
which stimulate further experimental and theoretical investigations
on the nature inherent in the Kitaev physics~\cite{
Sandilands2015,banerjee2016proximate,Banerjee2017,Do217,hirobe2017,Leahy2017,Hentrich2017pre,Kasahara2017pre,
Mazin2012,Knolle2014,Knolle2014raman,Nasu2014,Nasu2015,Zschocke2015,KimHS2015,Vojta2016,Nasu2016nphys,Rachel2016,Yoshitake2016,Yoshitake2017,Yoshitake2017b,Nasu2017,Metavitsiadis2017,Samarakoon2017}, and the generalization of the Kitaev model~\cite{Fendley2012,Vaezi2014,Barkeshli2015}.

Nevertheless, it is also recognized that
the magnetic properties at low temperatures
in the candidate materials cannot be fully reproduced by
the two-dimensional Kitaev model
while this model should capture the magnetism at higher temperatures.
For example, the materials exhibit a long-range magnetic order
at low temperatures~\cite{Singh2010,Singh2012,Plumb2014,Kubota2015,Sears2015,Majumder2015} and star-shape low-energy structure has been observed
by the inelastic neutron scattering experiments~\cite{Banerjee2017,Do217}.
To account for these features, a lot of theoretical investigations have been made for additional effects beyond the pure Kitaev model, such as Heisenberg and/or $\Gamma$ terms with/without long-range
interactions~\cite{Chaloupka2010,Jiang2011,Kimchi2011,Reuther2011,Chaloupka2013,Foyevtsova2013,Katukuri2014,Yamaji2014,Rau2014,Nasu2014Cu,Shinjo2015,Suzuki2015,Chaloupka2016,Yamaji2016,Sizyuk2016,Song2016,Okubo2017,winter2017,Catuneanu2017pre}
and intermediate spin-orbit coupling~\cite{Nakauchi}.
The previous works suggest that the Kitaev QSL survives
even in the small additional interactions owing to the existence of the spin gap
in the Kitaev model~\cite{Chaloupka2010,Chaloupka2013,Rau2014}, and the magnetic order observed in the real materials
is stable by the additional interactions within a two-dimensional honeycomb layer
in a plausible parameter range~\cite{Okubo2017}.
On the other hand, the candidate materials
are composed of honeycomb layers,
and the role of the stacking structure
in $\alpha$-RuCl$_3$ and H$_3$LiIr$_2$O$_6$
has been examined recently~\cite{Cao2016,Bette2017,Slagle2017pre}.
Therefore, it is highly desired to study the stability of the Kitaev QSL
in the presence of an interlayer coupling.

In this paper, we investigate the bilayer Kitaev model,
where two honeycomb layers are coupled by the Heisenberg interaction,
as one of simple models [see Fig.~\ref{fig:model}(a)].
We show that this model possesses a local conserved quantity
on each pair of stacked two hexagons (bi-hexagon),
which leads to the absence of the long-range magnetic order.
To study the stability of the Kitaev QSL against the interlayer coupling,
we employ the bond-operator mean-field (MF)
approximation~\cite{PhysRevB.41.9323} and
cluster expansion technique~\cite{SE1,SE2,SE3}.
We also use the exact diagonalization (ED) on finite clusters
to discuss ground state properties and excitation spectrum.
The numerical results suggest the existence of
a first-order phase transition between the Kitaev QSL and
spin-singlet dimer states at zero temperature.
Moreover, using the thermal pure quantum (TPQ) state
approach~\cite{Sugiura2012,Sugiura2013},
we clarity that the double-peak structure emerges
in the small interlayer coupling and retains
even above the quantum transition point expected from the ED calculations.
We also extend the argument based on the presence of
the local conserved quantities to the multilayer Kitaev model
and show the absence of three-dimensional long-range correlations in the multilayer Kitaev model with arbitrary stacking numbers as well.

This paper is organized as follows.
In Sec.~\ref{sec:model}, we introduce
the model Hamiltonian on the bilayer honeycomb lattice
and discuss the local conserved quantity and
the parity symmetry in the system.
Our methods are briefly summarized in Sec.~\ref{sec:method}.
In Sec.~\ref{sec:ground-state-prop},
we show the numerical results for the ground state properties in the bilayer system, which suggest the existence of a first-order quantum phase transition
between the QSL and singlet dimer states.
Thermodynamic properties are discussed in Sec.~\ref{sec:finite-temp-prop}.
In Sec.~\ref{sec:discussion},
we discuss magnetic properties in the multilayer Kitaev model.
The summary is provided in the last section.

\section{Model and its symmetry}\label{sec:model}

\begin{figure}
\centering
\includegraphics[width=\columnwidth]{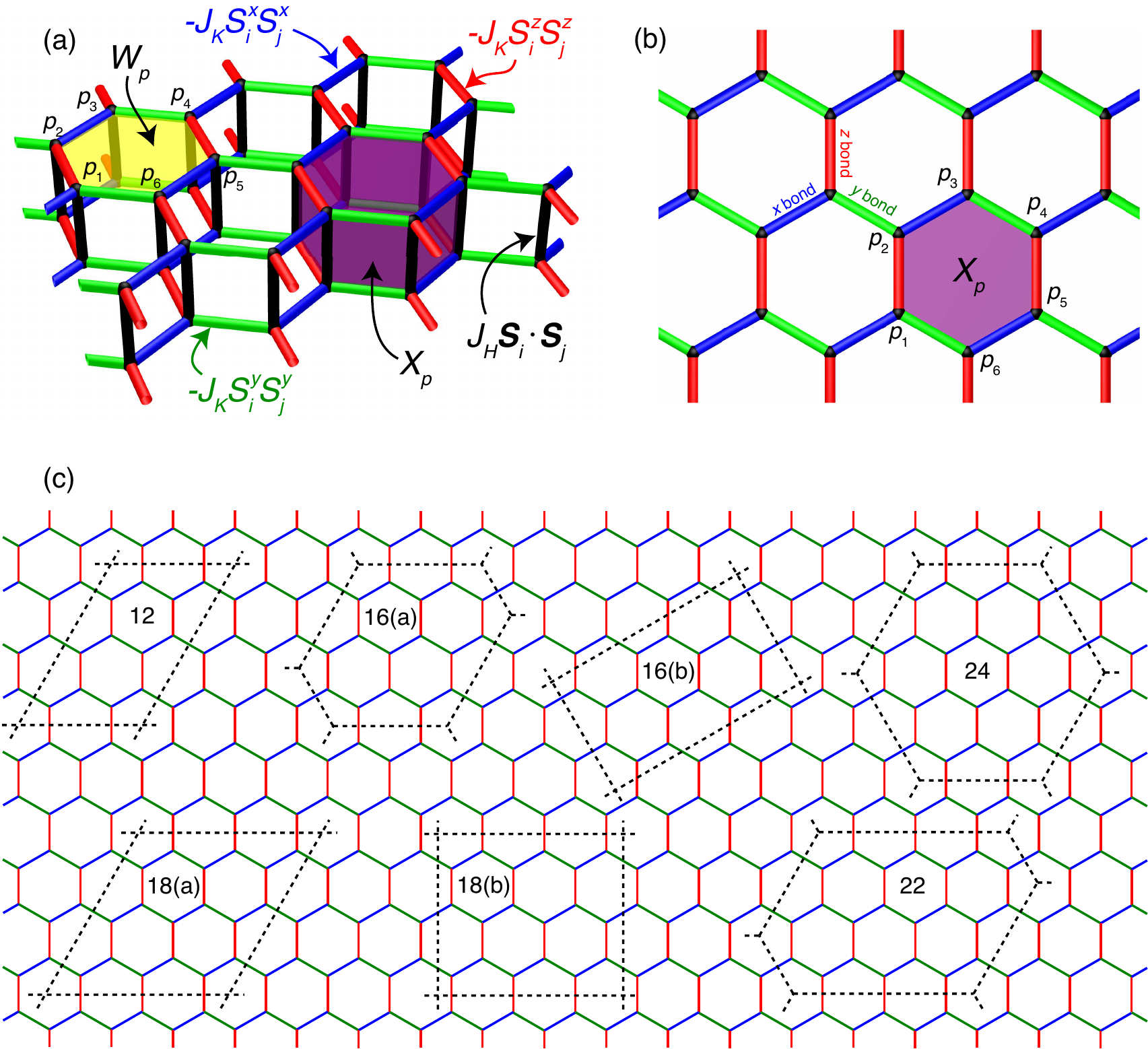}
\caption{
(a) Bilayer Kitaev model on the honeycomb lattice.
(b) Top view of the bilayer Kitaev model with the definition of dimer sites $p_i$ for the local conserved quantity $X_p$.
(c) Several clusters used in the ED calculations.
}
\label{fig:model}
\end{figure}

To address the effect of the interlayer coupling between Kitaev models
on honeycomb lattices, we introduce a following simple model,
where two Kitaev models are coupled by the Heisenberg interaction
[see Fig.~\ref{fig:model}(a)]:
\begin{align}
 {\cal H} = -  J_K\sum_{\means{ij}_\alpha,n} S_{i,n}^\alpha S_{j,n}^\alpha
 + J_H \sum_{i} {\bf S}_{i,1}\cdot{\bf S}_{i,2},\label{eq:1}
\end{align}
where $S_{i,n}^\alpha=\frac{1}{2}\sigma_{i,n}^\alpha (\alpha=x,y,z)$ and
$\sigma_{i,n}^\alpha$ is Pauli matrix at site $i$ of the $n(=1,2)$th layer.
$J_K (>0)$ is the ferromagnetic Kitaev coupling in each layer
and $J_H (>0)$ is the antiferromagnetic Heisenberg coupling between two layers.
We assume that each site on the layer 1 is located just above that
on the other.
In each layer, the anisotropy of the Ising-type interactions
depend on the bonds;
there are three kinds of nearest neighbor (NN) bonds,
${\langle ij \rangle}_\alpha$ ($\alpha=x,y,z$),
which we refer to as the $\alpha$ bond, on the honeycomb lattice
[see Figs.~\ref{fig:model}(a) and~\ref{fig:model}(b)].

When $J_H=0$, the system is reduced to two single-layer Kitaev models.
In the model, there exists the local $Z_2$ conserved quantity
$W_{p,n}=\sigma_{p_1,n}^x \sigma_{p_2,n}^y \sigma_{p_3,n}^z
\sigma_{p_4,n}^x \sigma_{p_5,n}^y \sigma_{p_6,n}^z$ on the $n$th layer ,
which results in the QSL ground state with long-range spin entanglement and the fractionalization of the quantum spins.
This allows us to map the single layer model onto free Majorana fermion system
with gapless elementary excitations
although there exists a gap in the spin excitation.
At finite temperatures, the spin fractionalization emerges as a peculiar temperature dependence of observables such as double-peak structure in the specific heat~\cite{Nasu2014,Nasu2015}.
On the other hand, in the case of $J_K=0$,
the system is composed of independent dimers.
The ground state is represented by the direct product of
interlayer dimer singlets with the spin gap, where inter-dimer wave functions are disentangled.
Although these two nonmagnetic ground states possess a spin gap,
their low-energy properties are different from each other.
Then, one naively expects a phase transition(s) between these two states
by changing the parameter $\lambda = J_H/J_K$.

The striking feature of this model is that
there exists a local conserved quantity.
In the presence of the interlayer coupling $J_H$, $W_p$ no longer commutes with the Hamiltonian, but the product $X_p=W_{p,1}W_{p,2}$ remains a local $Z_2$ conserved quantity [see Fig.~\ref{fig:model}(b)].
This is because two spin operators with dimer site $i$ in $X_p$
on plaquette $p$ have the same spin component,
and $X_p$ commutes with ${\bf S}_{i,1}\cdot{\bf S}_{i,2}$.
Therefore, the eigenstates of the Hamiltonian given in Eq.~(\ref{eq:1}) is characterized by the eigenvalue of $X_p$, $\pm1$, in each bi-hexagon $p$.
The existence of the local conserved quantity leads to the absence of spin correlations except for NN sites in Kitaev layers and interlayer dimers.
Then, we can say that the ground state is always nonmagnetic
even in the presence of interlayer coupling $J_H$.

To clarify another symmetry of the present bilayer Kitaev system given in Eq.~(\ref{eq:1}), we adopt the bond-operator representation~\cite{PhysRevB.41.9323},
which is useful for spin-dimer
systems~\cite{Gopalan1994,Kotov1998,Matsushita1999}.
In this representation, the four local bases on each dimer $i$ are taken as
\begin{align}
 \kets{s}_i&=s_i^\dagger \kets{0}=\frac{1}{\sqrt{2}}\left(\kets{\uparrow\downarrow}_i-\kets{\downarrow\uparrow}_i\right),\\
 \kets{t_x}_i&=t_{xi}^\dagger \kets{0}=-\frac{1}{\sqrt{2}}\left(\kets{\uparrow\uparrow}_i-\kets{\downarrow\downarrow}_i\right),\\
 \kets{t_y}_i&=t_{yi}^\dagger \kets{0}=\frac{i}{\sqrt{2}}\left(\kets{\uparrow\uparrow}_i+\kets{\downarrow\downarrow}_i\right),\\
 \kets{t_z}_i&=t_{zi}^\dagger \kets{0}=\frac{1}{\sqrt{2}}\left(\kets{\uparrow\downarrow}_i+\kets{\downarrow\uparrow}_i\right),
\end{align}
where $s_i^\dagger$ and $t_{\alpha i}^\dagger$ ($\alpha=x,y,z$) are the creation operators of the singlet and triplets on the dimer site $i$, respectively, and $\kets{0}$ is their vacuum.
We assume that these bond operators
behave as bosons and impose the local constraint $s_i^\dagger s_i +\sum_\alpha t_{\alpha i}^\dagger t_{\alpha i}=1$ on each dimer $i$ so as to reproduce the commutation relation of an $S=1/2$ spin.
By means of the bond operators, the Hamiltonian given in Eq.~(\ref{eq:1}) is rewritten as
\begin{align}
 {\cal H}=&-\frac{J_K}{2}\sum_{\means{ij}_\alpha}\Bigg(s_i s_j^\dagger t_{\alpha i}^\dagger t_{\alpha j} + s_i s_j t_{\alpha i}^\dagger t_{\alpha j}^\dagger + {\rm H.c.}\notag\\
&\qquad\qquad
-\sum_{\beta\beta'\gamma\gamma'}\epsilon_{\alpha\beta\gamma}\epsilon_{\alpha\beta'\gamma'}t_{\beta i}^\dagger t_{\gamma i}t_{\beta' j}^\dagger t_{\gamma' j}
 \Bigg)\notag \\
&+J_H\sum_i\left(-\frac{3}{4}s_i^\dagger s_i+\frac{1}{4}\sum_\alpha t_{\alpha i}^\dagger t_{\alpha i}\right)\notag \\
&-\sum_i\mu_i\left(s_i^\dagger s_i +\sum_\alpha t_{\alpha i}^\dagger t_{\alpha i}-1\right).\label{eq:2}
\end{align}
where $\epsilon_{\alpha\beta\gamma}$ stands for the Levi-Civita symbol and $\mu_i$ is a Lagrange multiplier on each dimer to impose the local constraint.

We find that the number of each boson is not conserved, but
each boson are created or annihilated as a pair.
For example, the second line in Eq.~(\ref{eq:2})
is expanded as
$t_{iy}^\dagger t_{iz} t_{jz}^\dagger t_{jy}-t_{iy}^\dagger t_{iz} t_{jy}^\dagger t_{jz}+{\rm H.c.}$
when $\alpha=x$.
Therefore, the parity of the particle number is conserved
in each component of bosons,
In other words, the parity operators,
$P_s=\exp[i\pi \sum_is_i^\dagger s_i]$ and
$P_{t_\alpha}=\exp[i\pi \sum_ it_{i\alpha}^\dagger t_{i\alpha}]$
($\alpha=x,y,z$) commute with the Hamiltonian.

\begin{figure}
\centering
\includegraphics[width=\columnwidth]{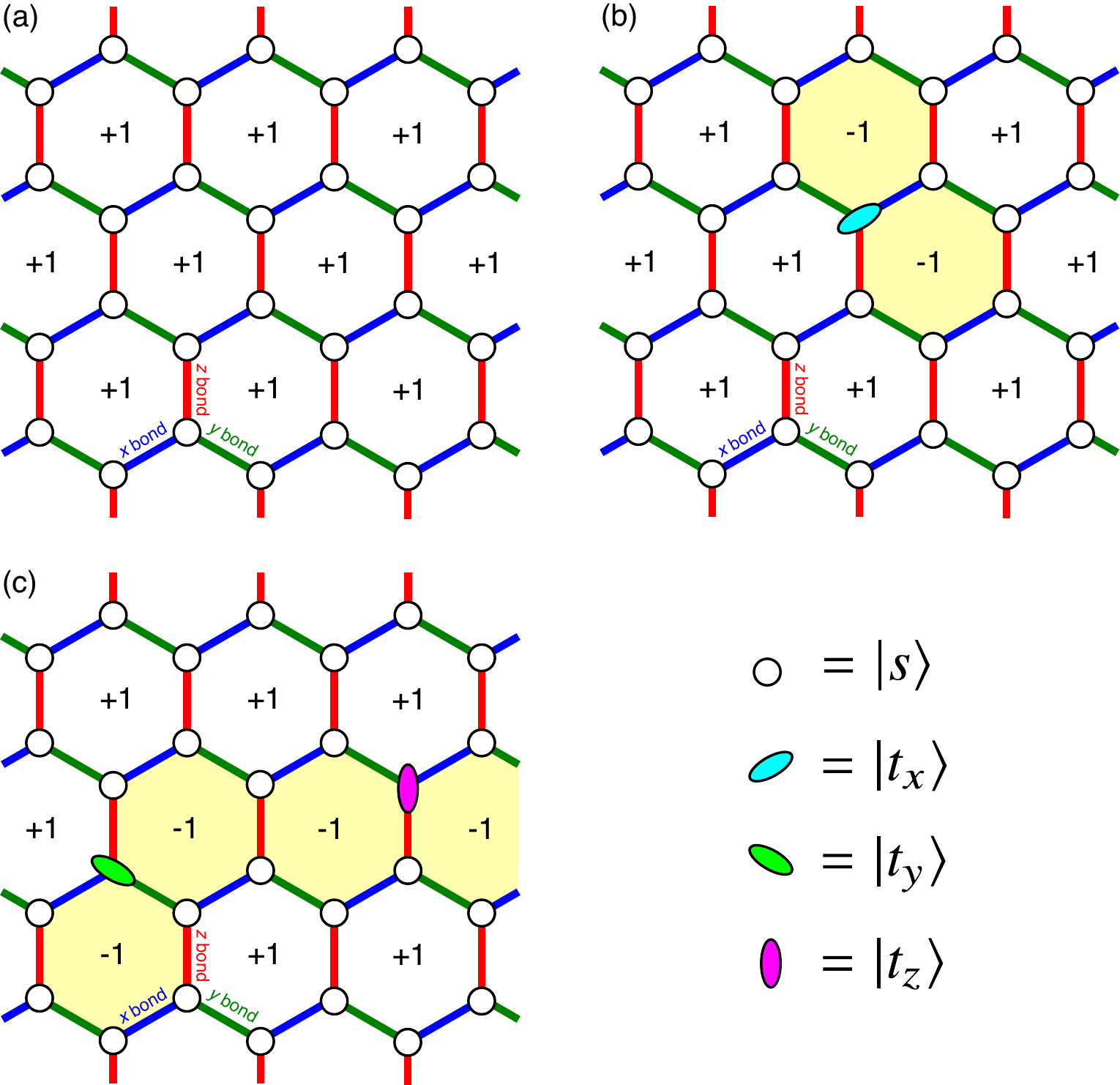}
\caption{
Configurations of eigenvalues of $X_p$ in (a) the state $\kets{\Phi_s}$ where all dimers are spin singlets, (b) one-triplon state $\kets{\Phi_{t_{xi}}}$ where an $x$ component of triplon is excited, and (c) two-triplon state $\kets{\Phi_{t_{yi},t_{zj}}}$ where $y$ and $z$ components of triplons are excited.
In these figures, open circles represent spin singlets, and colored ellipses stand for the three different components of spin triplets as shown in the lower right.
The shaded hexagons represent the plaquettes with $X_p=-1$.
}
\label{fig:xconfig}
\end{figure}

We note that a direct product of the four local states on a dimer,
$\kets{s}_i$, $\kets{t_x}_i$, $\kets{t_y}_i$, and $\kets{t_z}_i$ is
the eigenstate of $X_p$
since
\begin{eqnarray}
\sigma_{i1}^\alpha\sigma_{i2}^\alpha\kets{s}_i=-\kets{s}_i,\quad \sigma_{i1}^\alpha\sigma_{i2}^\alpha\kets{t_\beta}_i=
(1-2\delta_{\alpha\beta})\kets{t_\beta}_i.
\end{eqnarray}
In addition, the number operators of the bond operators are given as projectors onto the singlet and triplet states:
\begin{eqnarray}
 s_i^\dagger s_i=\frac{1}{4}-{\bf S}_{i1}\cdot{\bf S}_{i2},
\quad t_{\alpha i}^\dagger  t_{\alpha i} =\frac{1}{4}+{\bf S}_{i1}\cdot{\bf S}_{i2}-2S_{i1}^\alpha S_{i2}^\alpha,
\end{eqnarray}
leading to the fact that each parity operator commutes with $X_p$.
Therefore, the Hilbert space of the Hamiltonian can be classified into
each subspace specified by $[P_{t_x}, P_{t_y}, P_{t_z}, \{X_p\}]$, where
$P_s$ is determined by $P_{t_\alpha}$ and the number of dimers $N$
due to the local constraint.
For example, for the state $\kets{\Phi_s}=\prod_i s_i^\dagger \kets{0}$,
$P_{t_x}=P_{t_y}=P_{t_z}=+1$ and $X_p=+1$ for all plaquettes,
which is schematically shown in Fig.~\ref{fig:xconfig}(a).
Here, we can consider a state $\kets{\Phi_{t_{xi}}}=t_{ix}^\dagger
\prod_{j\neq i}s_j^\dagger\kets{0}=\tilde{t}_{ix}^\dagger \kets{\Phi_s}$,
where we have introduced the triplet excitation (triplon) creation operator
from the singlet state,
$\tilde{t}_{i\alpha}^\dagger = t_{i\alpha}^\dagger s_i$.
This state is also an eigenstate of $X_p$ with the eigenvalues of $-1$
for the adjacent two plaquettes with the shared $x$ bond,
where the triplon is sit on one of the edge sites.
[see Fig.~\ref{fig:model}(b)].
In the case, the eigenvalue of $X_{p'}$ is $+1$ for the other plaquettes $p'$.
From the above discussion,
one can determine the spatial configuration of eigenvalues of $X_p$,
as shown in Fig.~\ref{fig:xconfig}(b).
The eigenvalues of $X_p$ for the multiple triplet-excited state can be also discussed in the same manner, as shown in Fig.~\ref{fig:xconfig}(c).
Note that this is similar to the configuration of $W_p$
for the state where a spin is operated to the ground state
in the single-layer Kitaev model.
The details of the excitation spectrum of triplons will be given
in Sec.~\ref{sec:excitation-spectra}.

\section{Methods}\label{sec:method}

To analyze the properties of the ground state in Eq.~(\ref{eq:1}),
we mainly employ the ED method in finite size clusters.
The clusters used in the present calculations are
shown in Fig.~\ref{fig:model}(c).
In the ED calculations, we utilize the presence of the conserved quantities,
$X_p$ and the parities $P_s$, $P_{t_x}$, $P_{t_x}$, and $P_{t_z}$,
to reduce the matrix dimensions.
In addition to the ED calculations,
we also make use of the bond-operator MF theory~\cite{PhysRevB.41.9323} and
cluster expansion technique~\cite{SE1,SE2,SE3}
to perform the comprehensive analysis.
In the bond-operator MF theory, the local constraint is changed to the global one by introducing the uniform chemical potential $\mu=\mu_i$ and the Bose condensation of singlets are assumed as $\means{s_i}=\means{s_i^\dagger}=\bar{s}$~\cite{Gopalan1994,Kotov1998,Matsushita1999}.
Owing to these assumptions, the Hamiltonian is reduced to the free boson system of the triplet excitations as
\begin{align}
 {\cal H}_{\rm MF}=&\left(-\frac{3}{4}J_H\bar{s}^2-\mu\bar{s}^2+\mu\right)N+\left(\frac{J_H}{4}-\mu\right)\sum_{i,\alpha} t_{\alpha i}^\dagger t_{\alpha i}\notag\\
&-\frac{J_K\bar{s}^2}{2}\sum_{\means{ij}_\alpha}\left( t_{\alpha i}^\dagger t_{\alpha j} +  t_{\alpha i}^\dagger t_{\alpha j}^\dagger +{\rm H.c.}\right),\label{eq:3}
\end{align}
where we neglect the scattering terms between triplet excitations given in the second line of Eq.~(\ref{eq:2}), and $\mu$ and $\bar{s}$ are determined self-consistently.

We also use the cluster expansion technique~\cite{SE1,SE2,SE3},
which is one of the powerful methods to discuss the quantum phase transitions
in the frustrated quantum spin systems such as
$J_1$-$J_2$~\cite{Gelfand1989,PhysRevB.63.104420},
orthogonal-dimer~\cite{Weihong1999,PhysRevLett.84.4461,Koga2000b,PhysRevLett.85.3958,Takushima}
and Kitaev-Heisenberg models~\cite{PhysRevB.92.020405,PhysRevB.96.144414}.
In the bilayer model,
we begin with interlayer dimer singlets~\cite{Hida,Hida1992}.
As discussed before, there exists the parity symmetry
for the number of singlets and triplets in our model, and thereby
the odd-order coefficients never appear
in the ground state energy $E_\lambda/N$
and spin correlations $\langle {\bf S}_{i,1}\cdot {\bf S}_{i,2}\rangle$.
We compute the power series up to the 30th order for the above quantities.
Furthermore,
exploiting the first-order inhomogeneous differential method~\cite{Guttmann},
we deduce the quantities far from the dimer limit $\lambda\rightarrow\infty$.

To examine thermodynamic quantities at finite temperatures,
we employ the thermal pure quantum (TPQ) state
approach~\cite{Sugiura2012,Sugiura2013}.
In the paper, we treat the cluster with $N=12$ [see Fig.~\ref{fig:model}(c)].
We prepare more than 20 random vectors for initial states, and the physical quantities are calculated by averaging the values generated by these initial states.

\section{Numerical results}\label{sec:ground-state-prop}

\subsection{Ground State Properties}
\begin{figure}
\centering
\includegraphics[width=\columnwidth]{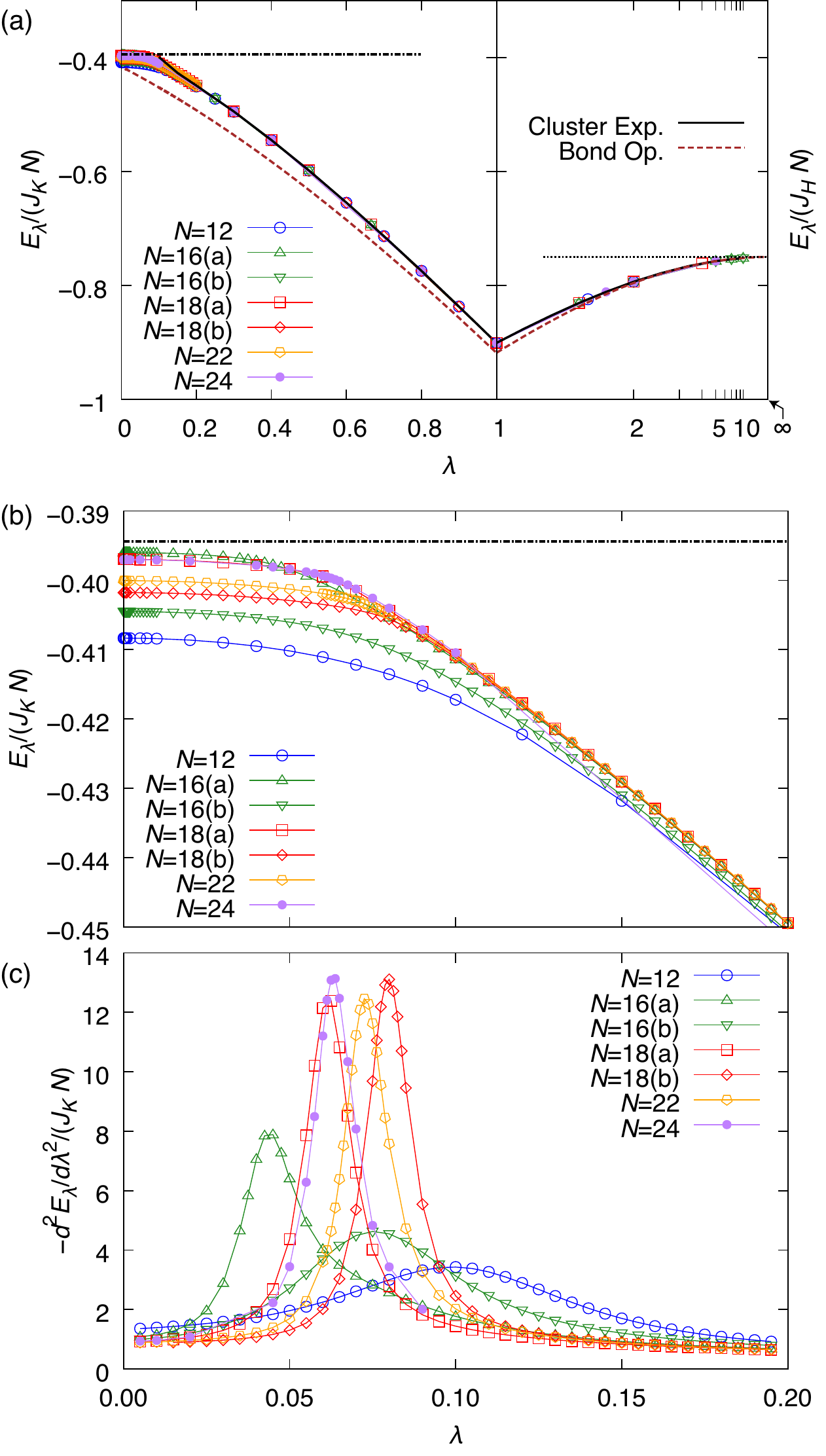}
\caption{
(a) Ground state energy per dimer as a function of $\lambda$
in the bilayer Kitaev model.
The solid and dashed lines represent the results obtained by the cluster expansion
and bond operator MF theory.
The symbols are obtained by the finite-size ED calculations.
The dashed-dotted and dotted lines represent the energies of the Kitaev QSL and singlet dimer states, respectively.
(b) Magnified plot of the small $\lambda$ region in (a).
(c) Second-order differential of the minimum energy per dimer.
}
\label{fig:emin}
\end{figure}

In this section, we discuss ground state properties
of the bilayer Kitaev model given in Eq.~(\ref{eq:1}).
Figure~\ref{fig:emin}(a) shows the $\lambda$ dependence of
the ground-state energy $E_\lambda$.
We obtain the smooth curve by the ED calculations.
We find that the energy obtained by the bond-operator MF approximation
is lower than that of the ED.
This should be due to its artifacts originating from the fact that
triplet-triplet correlations are not taken into account correctly.
With increasing $\lambda$, the difference between the ED and bond-operator
MF results becomes negligible,
meaning that the bond-operator MF approximation is justified
in the large $\lambda$ region.
We also show the energy obtained by the cluster expansion method
as the solid line in Fig.~\ref{fig:emin}(a).
Surprisingly, this almost coincides with the ED results
except for $\lambda\lesssim 0.2$ though the method is also an approach
from the large $\lambda$ limit.
This indicates that the dimer singlet state,
which is adiabatically connected to the direct product state $\kets{\Phi_s}$,
is realized in the region.

In the small $\lambda$ region, a large size dependence appears
in the ED results, as shown in Fig.~\ref{fig:emin}(b),
in contrast to the large $\lambda$ region.
With increasing the system size, the energy in the limit of $\lambda=0$ should
approach the exact value although it depends on the cluster shape.
An important point is that
the energy curve tends to have the bend structure around $\lambda\sim 0.06$.

The peculiar $\lambda$ dependence of the ground-state energy
can be clearly seen by taking its derivative.
Figure~\ref{fig:emin}(c) shows the second derivative of $E_\lambda$
as a function of $\lambda$.
It is found that the peak structure in each cluster
develops with increasing the system size
while the dependence of the cluster shape is also observed.
This indicates the existence of a first-order quantum phase transition
at $\lambda=\lambda_c(\sim 0.06)$ in the thermodynamic limit.

\begin{figure}
\centering
\includegraphics[width=\columnwidth]{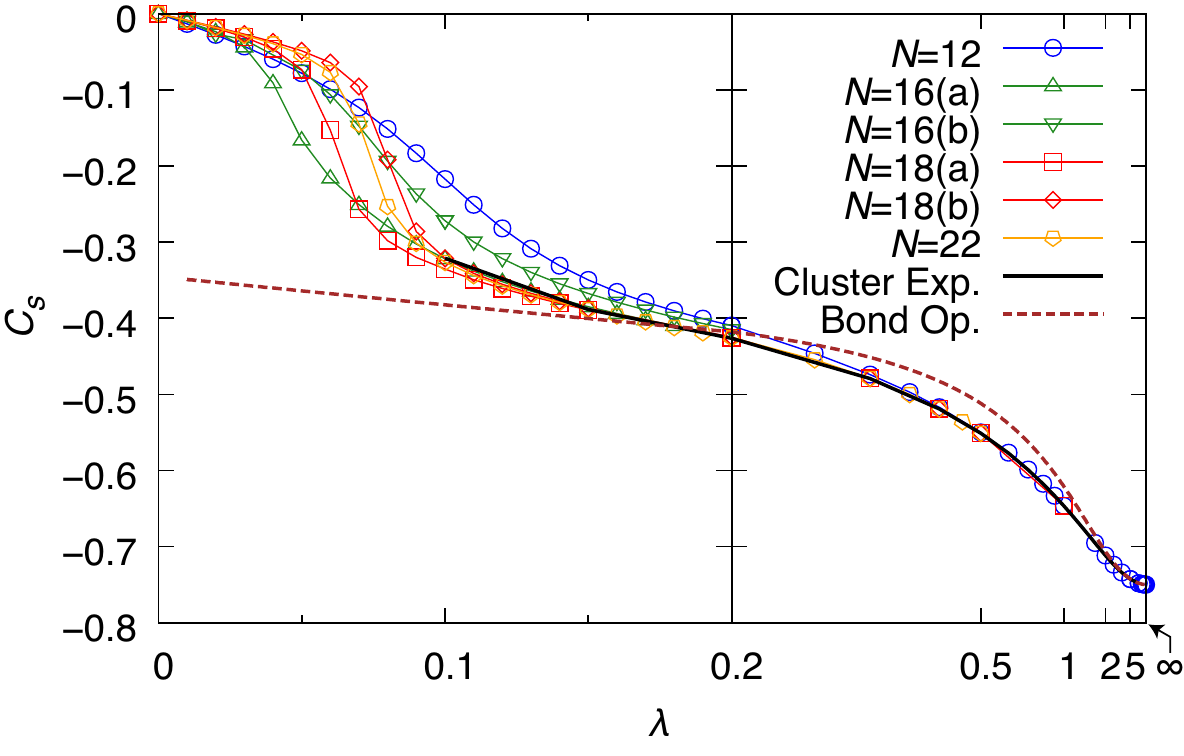}
\caption{
Interlayer dimer spin correlation $C_s$ as a function of $\lambda$.
The solid and dashed lines represent the results obtained by the cluster expansion and bond operator MF theory, and the symbols are obtained by the finite-size ED calculations.
}
\label{fig:SS}
\end{figure}

To clarify the nature of two distinct phases around $\lambda_c$,
we also calculate the interlayer spin correlations
$C_s=\frac{1}{N}\sum_i \means{{\bf S}_{i,1}\cdot {\bf S}_{i,2}}$.
The results are shown in Fig.~\ref{fig:SS}.
When $\lambda\to \infty$, the singlet dimer is realized in each site and
$C_s=-3/4$.
In the large $\lambda$ region, the strong interlayer coupling stabilizes
the dimer singlet state with a short correlation length.
Therefore, the ED results little depend on the cluster size and are
in a good agreement with those obtained by the cluster expansion,
implying that the dimer singlet state is realized.
With decreasing $\lambda$, the Ising coupling $J_K$ suppresses
interlayer spin correlations and thereby
the absolute value of $C_s$ decreases monotonically.
Around $\lambda=\lambda_c$, large cluster-size and shape dependence appears
in the ED results and it is hard to extrapolate the quantity
in the thermodynamic limit.
Nevertheless, the results for the larger clusters are still consistent
with the cluster expansion, which implies that the dimer-singlet state
is realized
when $\lambda>\lambda_c$.

On the other hand, in the smaller $\lambda$ region,
the spin correlation is suddenly changed and take a smaller value,
suggesting the realization of the Kitaev QSL expected
in the two independent Kitaev models.
It is worth noting that $\lambda_c=(J_H/J_K)_c\sim 0.06$ is close
to the spin gap of the Kitaev model, $\Delta_{K}/J_K\sim 0.065$.
This implies that the Kitaev QSL against the interlayer Heisenberg coupling is maintained by the existence of the spin gap inherent in the Kitaev model.

\subsection{Excitation spectrum}\label{sec:excitation-spectra}

\begin{figure}
\centering
\includegraphics[width=\columnwidth]{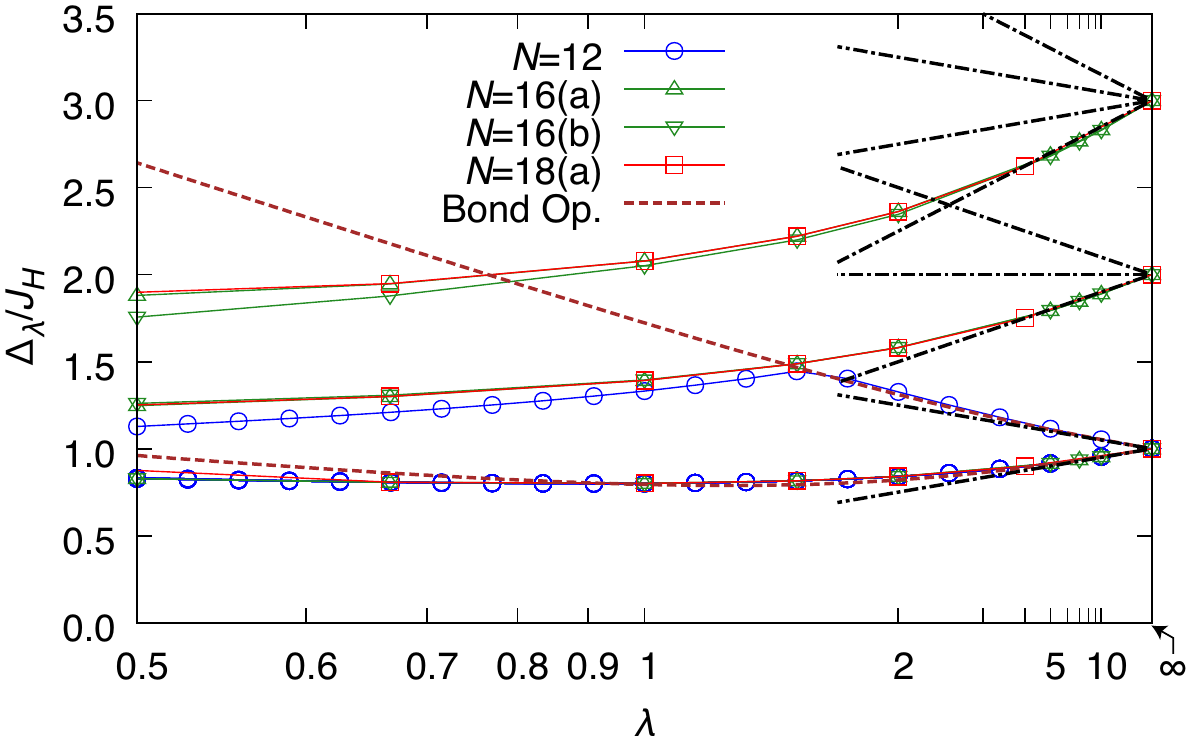}
\caption{
$\lambda$ dependence of the excitation energy $\Delta_\lambda$.
The symbols and dashed lines represent the results obtained by the ED method and bond operator MF theory, respectively.
For the $N=12$ cluster, excitation energies up to 19th excited states
are presented.
For the other clusters, excitation energies are calculated from
the lowest energy states of the following subspaces:
$(P_{t_x},P_{t_y},P_{t_z})=(-1,+1,+1)$, $(-1,-1,+1)$, and $(-1,-1,-1)$.
The dashed-dotted lines represent the excitation energies expected
from the localized bonding and antibonding energy levels,
$1\pm 1/(2\lambda)$, $2\pm 1/\lambda$, $2$, $3\pm 1/(2\lambda)$,
$3\pm 3/(2\lambda)$, in the large $\lambda$ limit.}
\label{fig:dispersion}
\end{figure}

Next, we discuss the excitation structure of the bilayer Kitaev model.
It is known that in the Kitaev model $(\lambda=0)$,
there is a Majorana continuum in the excited states.
On the other hand, in the dimer limit $\lambda\to \infty$,
due to the spin gap with the excitation energy $J_H$,
discrete excited levels appear, corresponding to the number of triplons.
Here, we first analyze the excitation spectrum by means of
the bond operator method.
The MF Hamiltonian Eq.~(\ref{eq:3}) indicates that
the $\alpha$ component of the local triplet excitation is hybridized
with that only on the NN $\alpha$ bond.
Thus, the triplet excitation $t_\alpha$ is localized on each $\alpha$ bond
and forms the bonding and antibonding states.
Figure~\ref{fig:dispersion} shows the one-body spectrum of the triplet excitation obtained by the bond operator MF approximation.
The introduction of the Kitaev interaction splits excitation energy of the spin gap into two levels, which corresponds to the formation of the bonding and antibonding states.

The fact that triplet excitations are localized on the corresponding NN bonds can be confirmed exactly by considering the spatial configuration of
the local conserved quantity $X_p$.
This is done in the similar manner as the proof for the absence of
the long-range spin correlations
in the single layer Kitaev model~\cite{Baskaran2007}.
As shown in Fig.~\ref{fig:xconfig}(b), in the one-triplon state $\kets{\Phi_{t_xi}}$ on dimer site $i$, the two eigenvalues of $X_p$ are flipped from the ground state.
On the other hand, one-triplon state $\kets{\Phi_{t_xj}}$
on another dimer site $j$ possesses
a different configuration of eigenvalues of $X_p$, and therefore
$\means{\Phi_{t_xi}|\Phi_{t_xj}}=\bras{\Phi_s}\tilde{t}_{xi} \tilde{t}_{xj}^\dagger\kets{\Phi_s}=0$ expect for the case where the pair $ij$ is on the NN $x$ bond.
As for one-triplon state with a different component, $\kets{\Phi_{t_\alpha j}}\; (\alpha\neq x)$,
its parity is different from that of the state $\kets{\Phi_{t_xj}}$
and these two states are never hybridized.
The above consideration leads to localized bonding and antibonding states for one triplon excitation,
which are $3N/2$-fold degenerate each.
The corresponding energies are
$\Delta_\pm/J_H=1\pm 1/(2\lambda) + 3/(8\lambda^2) + \cdots$.
In addition, one can considers the multiple excitations,
which are shown as the dashed-dotted lines in Fig.~\ref{fig:dispersion}.
Note that a pair of two triplons with the $\alpha$ component
on an $\alpha$ bond can be mixed with the singlet state $\kets{\Phi_s}$,
and hence, the bound state may have a dispersion
in the presence of the Kitaev interaction with $O(1/\lambda^2)$.
Furthermore, two triplon states with different components should form
the bound state.
Therefore, two triplon states should be split into
some dispersionless bound states and dispersive bands
on the introduction of the Kitaev coupling $J_K$.

To confirm the above consideration, we calculate the excitation spectrum using the ED method.
The results are presented in Fig.~\ref{fig:dispersion}.
The low-energy spectra obtained by the bond-operator MF approximation
is well reproduced by the finite-size calculation
in the large $\lambda$ case because of
the existence of the localized excitations.
We have also confirmed that the first and second excited states
in $\lambda\gtrsim 1.5$ are $3N/2$-fold degenerate,
which is adiabatically connected to $3N$-fold degenerate states
coming from local triplet excitations at $\lambda\rightarrow\infty$.
These degeneracies are consistent with those of the bonding and antibonding states of the triplet excitations discussed above.
At $\lambda\sim 1.5$, excitation energies originating
from one-triplon and two-triplon excited states intersect without mixing.
This is understood from the fact that the parity of the triplon number
is conserved in the present system,
which prohibits the mixing between one-triplon and two-triplon excited states.
We wish to note that the lowest excitation energy is always finite.
Furthermore, we can not find any tendency for closing the gap.
This is consistent with the fact that the singlet-dimer state is realized in the region $\lambda>\lambda_c$.

In the Kitaev limit, the excitation spectrum shows a continuum
coming from the Majorana fermions, as mentioned before.
On the other hand, low-energy one-triplon states are localized, whose excitation energy may be proportional to the interlayer spin correlation $C_s$.
We expect that, with decreasing $\lambda$,
the discrete levels disappear at
the first-order quantum phase transition point $\lambda_c$
and the Majorana continuum might appear below $\lambda_c$.
However, larger cluster calculations are needed
to clarify the spectral change around $\lambda_c$,
and this remains as a future work.

\section{Finite Temperature Properties}\label{sec:finite-temp-prop}

\begin{figure}
\centering
\includegraphics[width=\columnwidth]{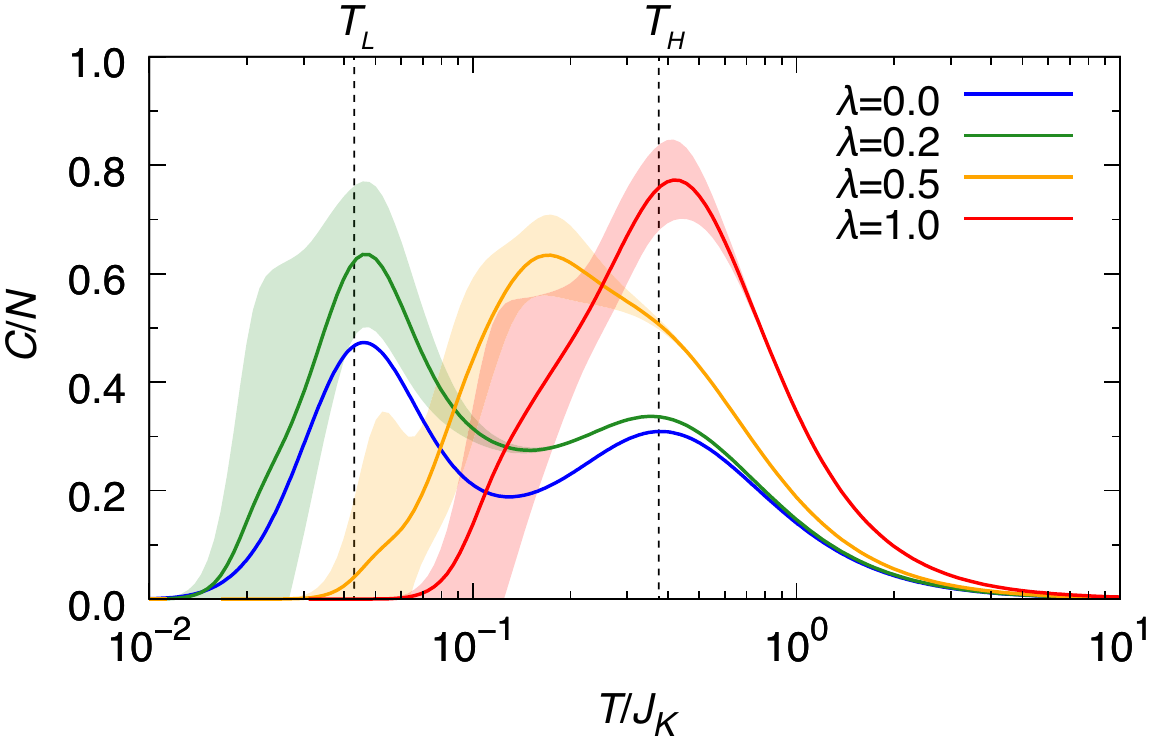}
\caption{
The specific heat per dimer $C/N$ as a function of the temperature
for the $N=12$ cluster when $\lambda=0, 0.1, 0.5$, and $1$.
Shaded areas are the possible errors estimated by using the standard deviation of the results
obtained from more than 20 initial random states in the TPQ calculations, and the result for $\lambda=0$ is obtained by the full diagonalization.
The dotted lines represent the peak temperatures at $\lambda=0$.
}
\label{fig:cv}
\end{figure}

Finally, we discuss finite temperature properties of the bilayer Kitaev model.
Figure~\ref{fig:cv} shows the temperature dependence of the specific heat
for several $\lambda$ by using the TPQ state for the $N=12$ cluster.
When $\lambda=0$, the system is reduced to the decoupled Kitaev models and
the double peak structure appears
at $T=T_L\sim 0.044J_K$ and $T_H\sim 0.38J_K$ in the $N=12$ cluster.
This is consistent with
the fact that the spin degrees of freedom is split into itinerant Majorana
fermions and localized gauge fluxes in each single-layer Kitaev model.
By comparing with the curves with $\lambda=0.2$,
the lower characteristic temperature $T_L$
slightly shifts to higher temperatures, while the other little shifts.
Further increasing $\lambda$,
these two peaks merge into a single peak,
which is a Schottky-type peak characteristic of the gapped systems.

\begin{figure}
\centering
\includegraphics[width=\columnwidth]{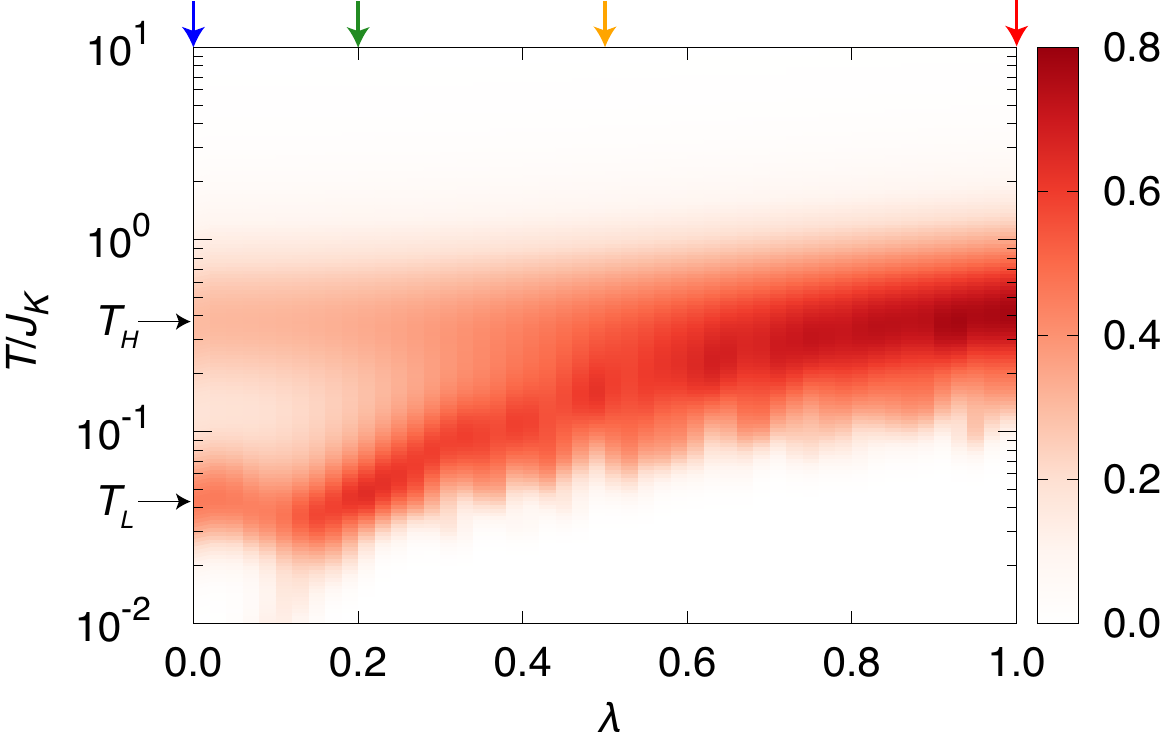}
\caption{
Density plot of the specific heat in the $T$-$\lambda$ plane.
The arrows represent the corresponding temperatures used in Fig.~\ref{fig:cv}.
}
\label{fig:tpq_c}
\end{figure}

To clarify how the double peak structure in the single-layer Kitaev model is gradually changed,
we show in Fig.~\ref{fig:tpq_c} the density plot of the specific heat
on the plane of $\lambda$ and temperature.
The double peaks exhibiting at $\lambda=0$ clearly appear
below $\lambda\sim 0.2$.
With increasing $\lambda$, the temperature of the higher-temperature peak hardly changes whereas the lower one depends on $\lambda$.
We find that these peaks merge into the single peak around $\lambda\sim 0.6$.
This suggests that the proximity effect of the Kitaev QSL emerges
at higher temperatures even above $\lambda_c$.
Similar effect has been discussed in the Kitaev-Heisenberg model,
where the double-peak structure is seen even in the magnetic ordered phase close to the Kitaev QSL ground state~\cite{Yamaji2016}.

\section{Discussion}\label{sec:discussion}

\begin{figure}
\centering
\includegraphics[width=\columnwidth]{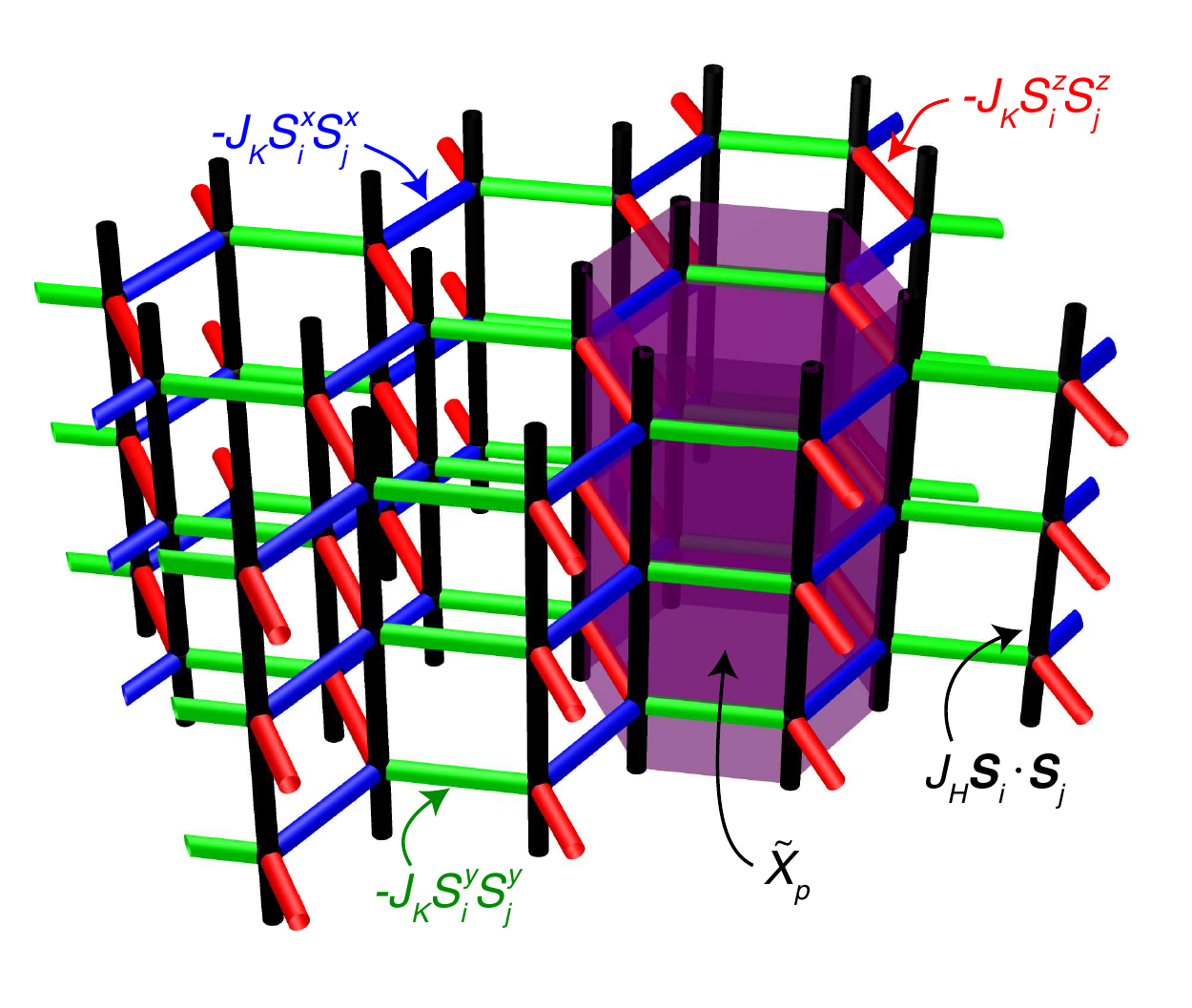}
\caption{
Schematic picture of the multilayer Kitaev model with interlayer Heisenberg couplings.
The local conserved quantity $\tilde{X}_p$ is depicted in the figure.
}
\label{fig:model3D}
\end{figure}

In this section, we briefly discuss magnetic properties of
the multilayer Kitaev systems,
where the Kitaev layers are connected by the Heisenberg couplings,
as shown in Fig.~\ref{fig:model3D}.
In the system, we can define the local $Z_2$ conserved quantity
$\tilde{X}_p=\prod_n W_{p,n}$ (see Fig.~\ref{fig:model3D}),
which is similar to the bilayer system discussed in the previous sections.
When the number of layers is large enough,
this system is regarded as one-dimensional Heisenberg chains
coupled with Kitaev interactions.
In the case, its spin correlation length shows power-law decay
along each chain direction.
However, the existence of the local conserved quantity $\tilde{X}_p$
indicates that there are no spin correlations
except for NN sites in Kitaev layers and sites
belonging to the same Heisenberg chain.
This suggests that the ground state of the stacked Kitaev model
with the interlayer Heisenberg couplings shown Fig.~\ref{fig:model3D}
is nonmagnetic as three-dimensional longer range spin correlations are absent.

\section{Summary}\label{sec5}

In summary, we have investigated the ground-state and the finite temperature
properties of the bilayer Kitaev model.
The results obtained by the ED, bond-operator MF,
and cluster expansion methods suggest
the existence of a first-order quantum phase transition
between the Kitaev QSL and singlet-dimer states in the thermodynamic limit.
In the excitation spectrum, the one-triplon excitation is localized
in the singlet-dimer phase,
which is proved by the existence of a local conserved quantity.
We have also discussed finite-temperature properties, where
the double-peak structure intrinsic to the Kitaev QSL appears
even in the singlet-dimer phase near the phase boundary.
Furthermore, we have shown that three-dimensional long-range
spin correlations are also absent in the stacked multilayer Kitaev systems
with arbitrary layer numbers.

\begin{acknowledgments}
The authors would like to thank S. Suga for valuable discussions.
This work is supported by Grant-in-Aid for Scientific Research from
JSPS, KAKENHI Grant Number JP17K05536, JP16H01066 (A.K.) and JP16K17747, JP16H02206, JP16H00987 (J.N.)
Parts of the numerical calculations were performed
in the supercomputing systems in ISSP, the University of Tokyo.
\end{acknowledgments}

\bibliography{./refs}

\end{document}